\documentstyle[12pt]{article}
\def\beqa{\begin{eqnarray}}
\def\eeqa{\end{eqnarray}}
\def\beq{\begin{equation}}
\def\eeq{\end{equation}}
\begin{document}
\begin{titlepage}
        \title{Inertial Effects on Neutrino Oscillations}
\author{S. Capozziello\thanks{E-mail: capozziello@physics.unisa.it}~~
and G. Lambiase\thanks{E-mail: lambiase@physics.unisa.it} \\ {\em Dipartimento di
Scienze Fisiche "E.R. Caianiello"} \\
 {\em Universit\'a di Salerno, 84081 Baronissi (Sa), Italy.} \\
 {\em Istituto Nazionale di Fisica Nucleare, Sez. di Napoli, Italy.} \\ }
\date{\today}
\maketitle
\begin{abstract}

The inertial effects on neutrino oscillations induced by the acceleration and angular
velocity of a reference frame are calculated. Such effects have been analyzed in the
framework of the solar and atmospheric neutrino problem.

\end{abstract}
\thispagestyle{empty} \vspace{20.mm} PACS number(s): 14.60.Pq, 95.30.Sf \\ \vspace{5.mm}
Keyword(s): Neutrino oscillations, Gravitational field \vfill
\end{titlepage}

\section{Introduction}
The long--standing problem about the deficiency of solar neutrino, i.e. a discrepancy
between the measured $\nu_{e}$ flux predicted by various solar models \cite{JNB},
\cite{BJN} and the atmospheric neutrino problem \cite{KSK}, might be explained invoking
oscillations between the various flavors or generations of neutrinos. It is well known,
in fact, that neutrino oscillations \cite{SMB} can occur, in the vacuum, if the
eigenvalues of the mass matrix are not all degenerate, and the corresponding mass
eigenstates are different from weak interaction eigenstates $\nu_{e}, \nu_{\mu},
\nu_{\tau}$. The most often discussed version of this type of solutions is the
Mikheyev--Smirnov--Wolfeinstein effect \cite{MSW} in which the solar electron neutrinos
can be converted, almost completely, into muon or tau neutrinos due to the presence of
matter in the Sun. Recently a quantum field theory of neutrino oscillations has been
proposed by Blasone, Vitiello \cite{VIT} and Sassaroli \cite{SAS}.

An alternative mechanism of neutrino oscillations, which does not require neutrino to
have a non--zero mass, was first suggested by Gasperini \cite{GAS} and by Halprin and
Leung \cite{HAL} as a means to test the equivalence principle. In this mechanism,
neutrino oscillations occur as a consequence of an assumed flavor non--diagonal coupling
of neutrinos to gravity which violates the equivalence principle. This line of research
has been followed also in Refs. \cite{ALL}.  A new solution of the solar neutrino
problem has been proposed in Ref. \cite{LIU} by using the mechanism introduced by Ellis,
Hagelin, Nanopoulos and Srednicki \cite{ELL}, in which the effect of quantum mechanics
violation, due to quantum gravity on neutrino oscillation, is investigated.

The effect of gravitationally induced quantum mechanical phases in neutrino oscillation
have been discussed in Ref. \cite{DVA}. Ahluwalia and Burgard consider the gravitational
effect on the neutrino oscillations showing that an external weak gravitational field of
a star adds a new contribution to the phase difference. They also suggest that the new
oscillation phase may be a significant effect on the supernova explosions since the
extremely large fluxes of neutrinos are produced with different energies corresponding
to the flavor states. This result has been critically riexamined by Bhattachya, Habib
and Mottola \cite{MOT}. They showed that the possible gravitational effect appears at
higher order with respect to that calculated in Ref. \cite{DVA}, with a magnitude of the
order $10^{-9}$, which is completely negligible in typical astrophysical applications.

Neutrino oscillations in curved space time have been also studied by Piniz, Roy, Wudka
\cite{PIN}, which observe that spin flavor resonant transitions of neutrinos may occur
in vicinity of active galactic nuclei due to gravitational effects and presence of a
large magnetic field, and by Cardall and Fuller \cite{CAR} which introduce an approach
to show that gravitational (like Schwarzschild field) effects on neutrino oscillations
are intimately related to the redshift.

The purpose of this paper is to calculate the contribution to neutrino oscillations
induced by inertial effects arising from acceleration and rotation of reference frames.
As it is well known, these effects are relevant in interferometry experiments. In fact,
by using an accelerated neutron interferometer, Bonse and Wroblewski were able to find
the predict phase shift \cite{BON}. Due to the validity of the equivalence principle,
one expects that this effect occurs also in a gravitational field, as verified by
Colella, Overhausen, Werner \cite{COL}. Besides, Mashhoon has derived a coupling of
neutron spin to the rotation of a non--inertial reference frame \cite{MAS} from an
extension of the hypothesis of locality, Atwood et al. found the neutron Sagnac effect
using an angular velocity of about $30$ times that of Earth \cite{ATW}, and finally,
Papini, Cai, Lloyd calculate the spin--rotation and spin--acceleration contributions to
the helicity precession of fermions \cite{PAP}.

At the present, there is a strong evidence in favour of oscillations of solar and
atmospheric neutrinos and of their non--zero masses. Such results have been found in
different experiments: 1) solar neutrino experiments \cite{HKE,KAM,GAL,SAG,SKD}, 2)
atmospheric neutrino experiments \cite{SKK,MIO,IMB,SOU,MAC}, and 3) the accelerator LSND
experiment \cite{LSND}. Nevertheless, we have to note that many other neutrino
oscillation experiments with neutrinos produced by reactors and accelerators did not
find any evidence of neutrino oscillations.

Recent reports indicate that the best fit in favour of neutrino oscillations are
obtained for the following cases \cite{ALH}: \\
 (MSW) small angle mixing region
 $$
 \vert m_2^2-m_1^2\vert \simeq (3\div 10)\cdot
 10^{-6}\mbox{eV$^2$}\,{,}\quad \sin^2 2\theta\simeq (0.6\div
 1.3)\cdot 10^{-2}\,{;}
 $$
 (MSW) large angle mixing region
 $$
 \vert m_2^2-m_1^2\vert \simeq (1\div 20)\cdot
 10^{-5}\mbox{eV$^2$}\,{,}\quad \sin^2 2\theta\simeq 0.5\div
 0.9\,{;}
 $$
 solar vacuum oscillation
 $$
 \vert m_2^2-m_1^2\vert \simeq (0.5\div 1.1)\cdot
 10^{-10}\mbox{eV$^2$}\,{,}\quad \sin^2 2\theta\simeq 0.67\div
 1\,{;}
 $$
 atmospheric neutrino oscillation (see also \cite{ALL1,GONZ})
 $$
 \vert m_2^2-m_1^2\vert \simeq (10^{-3}\div 10^{-2})
 \mbox{eV$^2$}\,{,}\quad \sin^2 2\theta\geq 0.8\,{,}
 $$
 $$
 \vert m_2^2-m_1^2\vert \simeq (0.5\div 6)\cdot
 10^{-3}\mbox{eV$^2$}\,{,}\quad \sin^2 2\theta \geq 0.82 \,{;}
 $$
 LSND experiment
 $$
 \vert m_2^2-m_1^2\vert \simeq (0.2\div 10)
 \mbox{eV$^2$}\,{,}\quad \sin^2 2\theta\simeq (0.2\div
 3)\cdot 10^{-2}\,{;}
 $$
$\vert m_2^2-m_1^2\vert$ is the mass--squared difference of neutrinos and $\theta$ the
mixing angle. In the following, we will restrict  to consider only the cases of solar
and atmospheric neutrino oscillations in the vacuum.

The layout of the paper is the following. In Sect. 2 we shortly discuss Dirac equation
in curved space--time and we calculate the probability that neutrino flavor oscillations
occur with respect to an accelerating and rotating observer. In Sect. 3 we discuss the
phenomenological consequences of inertial effects on solar and atmospheric neutrino
problem. Conclusions are drawn in Sect. 4.

\section{Neutrino Oscillations Induced by Accelerations and Rotations}

As in Ref. \cite{CAR}, the generalized neutrino phase is given by
\begin{equation}\label{1}
\vert\psi_{f}(\lambda)> = \sum_{j} U_{f j} e^{i\int_{\lambda_0}^{\lambda}P\cdot
p_{null}d\lambda^{\prime}}\vert\nu_j>\,{,}
\end{equation}
where $f$ is the flavor index and $j$ the mass one. $U_{f j}$ are the matrix elements
transforming flavor and mass bases, $P$ is the four--momentum operator generating
space--time translation of the eigenstates and $p^{\mu}_{null}=dx^{\mu}/d\lambda$ is the
tangent vector to the neutrino worldline $x^{\mu}$, parameterized by $\lambda$. The
covariant Dirac equation in curved space--time \cite{WEI} is
$[i\gamma^{\mu}(x)D_{\mu}-mc/\hbar ]\psi=0$, where the matrices $\gamma^{\mu}(x)$ are
related to the usual Dirac matrices $\gamma^{\hat{a}}$ by means of the vierbein fields
$e_{\mu}^{\hat{a}}(x)$, where the Greek (Latin with hat) indices refer to curved (flat)
space--time. $D_{\mu}$ is defined as $D_{\mu}=\nabla_{\mu}+\Gamma_{\mu}(x)$, where
$\nabla_{\mu}$ is the usual covariant derivative and $\Gamma_{\mu}(x)$ is the spinorial
connection defined by
 $$ \Gamma_{\mu}(x)={1\over
8}[\gamma^{\hat{a}}, \gamma^{\hat{b}}]e^{\nu}_{\hat{a}}e_{\nu \hat{b};\mu}\,{,}
 $$
 (semicolon represents the covariant derivative). The relations
 $$ \gamma^{\hat{a}}[\gamma^{\hat{b}},
\gamma^{\hat{c}}]=2\eta^{\hat{a}\hat{b}}\gamma^{\hat{c}}-2\eta^{\hat{a}\hat{c}}\gamma^{\hat{b}}
-2i\varepsilon^{\hat{d}\hat{a}\hat{b}\hat{c}}\gamma^5\gamma^{\hat{d}}\,{,}
 $$
 where $\eta^{\hat{a}\hat{b}}$ is the metric tensor in flat
spacetime, $\varepsilon^{\hat{d}\hat{a}\hat{b}\hat{c}}$ is the totally antisymmetric
tensor, $\gamma^5=i\gamma^{\hat{0}}\gamma^{\hat{1}}\gamma^{\hat{2}}\gamma^{\hat{3}}$ and
$\{\gamma^5, \gamma^{\hat{a}}\}=0$, allow to recast the non--vanishing contribution from
the spin connection in the form
 \begin{equation}\label{2}
\gamma^{\hat{a}}e^{\mu}_{\hat{a}}\Gamma_{\mu}=\gamma^{\hat{a}}e^{\mu}_{\hat{a}}
\left\{iA_{G\mu}\left[-(-g)^{-1/2}\frac{\gamma^5}{2}\right]\right\}\,{,}
 \end{equation}
where
 \begin{equation}\label{3}
A_G^{\mu}=\frac{1}{4}\sqrt{-g}e^{\mu}_{\hat{a}}\varepsilon^{\hat{d}\hat{a}\hat{b}\hat{c}}
(e_{\hat{b}\mu;\sigma}-e_{\hat{b}\sigma;\nu})e^{\nu}_{\hat{c}}e^{\sigma}_{\hat{d}}\,{,}
 \end{equation}
and $g\equiv det(g_{\mu\nu})$. $g_{\mu\nu}$ is the metric tensor of curved space--time.
The momentum operator $P_{\mu}$, used to calculate the phase of neutrino oscillations,
is derived from the mass shell condition
 \begin{equation}\label{4}
(P_{\mu}+\hbar A_{G\mu}\gamma^5)(P^{\mu}+\hbar A_G^{\mu}\gamma^5)=-M^2_f c^2\,{,}
 \end{equation}
 where $M^2_f$ is the vacuum mass matrix in flavor base
 \begin{equation}\label{5}
M^2_f=U\left(\begin{array}{cc}
                m_1^2 & 0 \\
                 0 & m_2^2 \end{array}\right)U^{\dagger}\,{,}\qquad
U =\left(\begin{array}{cc}
                \cos\theta & \sin\theta \\
                 -\sin\theta & \cos\theta \end{array}\right)\,{.}
 \end{equation}
$\theta $ is the vacuum mixing angle. Ignoring terms of the order ${\cal O}(\hbar^2
A_G^2)$ and ${\cal O}(\hbar A_GM_f)$, one gets that, for relativistic neutrinos moving
along generic trajectories parameterized by $\lambda$, the column vector of flavor
amplitude
 \begin{equation}\label{6}
\chi (\lambda)=\left(\begin{array}{c}
                           <\nu_e\vert \psi(\lambda)> \\
                           <\nu_\mu\vert \psi(\lambda)> \end{array}\right)
 \end{equation}
 satisfies the equation
 \begin{equation}\label{7}
i\frac{d\chi}{d\lambda}=\left(\frac{M_f^2c^2}{2}+\hbar p\cdot
A_G\gamma^5\right)\chi\,{.}
 \end{equation}
 In deriving Eq. (\ref{7}), one uses the relation $P^0=p^0$ and
$P^i\approx p^i$ \cite{CAR}. In an accelerating and rotating frame, the vierbein fields
$e_{\mu}^{\hat{a}}(x)$ are given by \cite{HEH}
 \begin{equation}\label{8}
e_0^{\hat{0}}=1+\frac{\vec{a}\cdot\vec{x}}{c^2}\,{,}\quad e_m^{\hat{0}}=0\,{,}\quad
e_0^{\hat{k}}=\varepsilon^{\hat{k}\hat{l}\hat{m}}\omega^{\hat{l}}x^{\hat{m}}\,{,}\quad
e_l^{\hat{k}}=\delta_l^k\,{,}
 \end{equation}
 where $k,l,m=1,2,3$, $x^{\mu}=(x^0, \vec{x})$ are the local coordinates
for the observer at the origin and $\vec{a}$, $\vec{\omega}$ are the acceleration and
angular velocity of the frame, respectively. The components $e^{\mu}_{\hat{a}}(x)$ and
$e_{\mu \hat{a}}(x)$ are calculated by using the metric tensors $g_{\mu\nu}$ and
$\eta_{\hat{a}\hat{b}}$, with $g_{\mu\nu}$ given by the element line \cite{HEH}
 \begin{equation}\label{9}
ds^2=\left[\left(1+\frac{\vec{a}\cdot\vec{x}}{c^2}\right)^2+
\left(\frac{\vec{\omega}\cdot\vec{x}}{c}\right)^2-
\frac{(\vec{\omega}\cdot\vec{\omega})(\vec{x}\cdot\vec{x})}{c^2}\right](dx^0)^2
-2dx^0d\vec{x}\cdot\frac{(\omega\land\vec{x})}{c}-d\vec{x}\cdot d\vec{x}\,{.}
 \end{equation}
Eq. (\ref{9}) has been derived by the following consideration (see \cite{HEH}): a local
inertial frame accelerates and rotates relative to Earth (we are considering Earth in
view of the following discussion where the detector of neutrinos is comoving with it),
owing to earth's gravity and rotation. Then, as a consequence, the local physics in this
frame is that of special relativity, provided that the effects induced by the curvature
can be neglected. In this meaning, an observer in a "stationary laboratory" on Earth
finds himself in a noninertial frame, and inertial effects arise due to the acceleration
and rotation.

 Inserting Eq. (\ref{8}) into (\ref{3}), one gets the components of $A_G^{\mu}$
 \begin{equation}\label{10}
A_G^0=0\,{,}\quad
\vec{A}_G=\frac{\sqrt{-g}}{2}\frac{1}{1+\frac{\vec{a}\cdot\vec{x}}{c^2}}
\left\{2\frac{\vec{\omega}}{c}-\frac{1}{c^2}[\vec{a}\land
(\vec{x}\land\vec{\omega})]\right\} \,{,}
 \end{equation}
so that Eq. (\ref{7}) becomes
 \begin{equation}\label{11}
i\frac{d}{d\lambda}\left(\begin{array}{c}
                           a_e \\
                           a_{\mu}\end{array}\right)={\cal T}\left(\begin{array}{c}
                                                            a_e \\
                                                          a_{\mu}\end{array}\right)\,,
 \end{equation}
 where $a_f\equiv <\nu_f\vert\psi(\lambda)>, f=e,\mu$ and the
matrix ${\cal T}$ is defined as
\begin{equation}\label{12}
{\cal T}=\left[\begin{array}{cc}
                -(\Delta/2)\cos2\theta  &   (\Delta/2)\sin 2\theta-\hbar\vec{p}\cdot\vec{A}_G \\
(\Delta/2)\sin 2\theta-\hbar\vec{p}\cdot\vec{A}_G  & (\Delta/2)\cos2\theta
\end{array}\right]\,,
\end{equation}
up to the $(m_1^2+m_2^2)c^2/2$ term, proportional to identity matrix. Here $\Delta\equiv
(m_2^2-m_1^2)c^2/2$. We restrict to flavors $e, \mu$, but this analysis works also for
different neutrino flavors. To determine the mass eigenstates $\vert\nu_1>$ and
$\vert\nu_2>$, corresponding to a fixed value of the acceleration and angular velocity
of the frame (i.e. for a fixed value of the affine parameter $\lambda$), one has to
diagonalize the matrix ${\cal T}$. Using the standard procedure, one writes the mass
eigenstates as a superposition of flavor eigenstates
 \begin{equation}\label{13}
\vert\nu_1(\lambda)>=\cos\tilde{\theta}(\lambda)\vert \nu_e>-\sin\tilde{\theta}
(\lambda)\vert \nu_{\mu}>\,{,}
 \end{equation}
$$ \vert\nu_2(\lambda)>=\sin\tilde{\theta}(\lambda)\vert \nu_e>+
\cos\tilde{\theta}(\lambda)\vert \nu_{\nu}>\,{,} $$
 where the mixing angle $\tilde{\theta}$ is defined in terms of the vacuum mixing angle
 \begin{equation}\label{14}
\tan 2\tilde{\theta}=\frac{\Delta\sin 2 \theta-2\hbar\vec{p}\cdot\vec{A}_G}{\Delta\cos
2\theta}\,{.}
 \end{equation}
 We note that $\tilde{\theta}\to \theta$ as $\vec{A}_G\to 0$
(i.e. $\vec{a}\to 0, \vec{\omega}\to 0$). The corresponding eigenvalues are
 \begin{equation}\label{15}
\tau_{1,2}=\pm\sqrt{\frac{\Delta^2}{4}\cos^2 2\theta +\left[\frac{\Delta}{2}\sin
2\theta-(\vec{p}\cdot\vec{A}_G)\right]^2 }\,{.}
\end{equation}
 Writing $\vert\psi(\lambda)>=a_1(\lambda)\vert\nu_1>+a_2(\lambda)\vert\nu_2>$, Eq. (\ref{11})
assumes the form
 \begin{equation}\label{16}
i\frac{d}{d\lambda}\left(\begin{array}{c}
                           a_1 \\
                          a_2\end{array}\right)=\left(\begin{array}{cc} \tau_1 & 0 \\
                                                   0 & \tau_2 \end{array}\right)                                                       \left(\begin{array}{c}
                                                            a_1 \\
                                                          a_2\end{array}\right)\,{,}
 \end{equation}
 where $a_i=<\nu_i\vert\psi(\lambda)>, i=1,2$, and
 \begin{equation}\label{17}
\left(\begin{array}{c}
            a_1 \\
           a_2   \end{array}\right)=\tilde{U}\left(\begin{array}{c}
                           a_e \\
                          a_{\mu}\end{array}\right)
\,{,}\quad  \tilde{U}=\left(\begin{array}{cc}
                 \cos\tilde{\theta} & \sin\tilde{\theta} \\
                  -\sin\tilde{\theta} &  \cos\tilde{\theta} \end{array}\right) \,{.}
 \end{equation}
 We used the condition $d\tilde{\theta}/d\lambda \approx 0$ in order that
(\ref{16}) is a diagonal matrix. It means that we are neglecting the variations of
acceleration and angular velocity, with respect to the affine parameter $\lambda$, in
comparing to their magnitudes. Eq. (\ref{16}) implies $a_i(\lambda)=a_i(0)\exp\alpha
(\lambda ), \alpha(\lambda)\equiv i\int_{\lambda_0}^{\lambda}\tau_i  d\lambda^{\prime},
i=1,2$. For the initial condition $\vert\psi (0)>=\vert\nu_e>$, the state $\vert\psi
(\lambda)>$ is
 \beq\label{18}
 \vert\psi (\lambda)>=[\cos\theta_0\cos\tilde{\theta}e^{i\alpha}+
 \sin\theta_0\sin\tilde{\theta}e^{-i\alpha}]\vert\nu_e>+
[-\cos\theta_0\sin\tilde{\theta}e^{i\alpha}+\sin\theta_0\cos\tilde{\theta}e^{-i\alpha}]
\vert\nu_{\mu}>\,{,}
 \eeq
 where $\theta_0=\tilde{\theta}(\lambda_0)$. The probability to observe
an electronic neutrino is therefore
 \beq\label{19}
 \vert <\nu_e\vert\psi(\lambda
)>\vert^2=\cos^2(\theta_0+\tilde{\theta})\sin^2\alpha +
\cos^2(\theta_0-\tilde{\theta})\cos^2\alpha\,{.}
 \eeq
 Eq. (\ref{19}) shows that accelerating and rotating observers will
experience a flavor oscillation of neutrinos. Due to the equivalence principle, one
concludes that gravitational fields can induce neutrino oscillations, in agreement with
Refs. \cite{GAS}--\cite{CAR}. It is interesting to discuss some particular case, i.e.
the frame is accelerating or rotating, in order to estimate the contributions to
neutrino oscillations when inertial effects are taken into account.

\section{Inertial Effects on Solar and Atmospheric Neutrinos}

Consequences on neutrino oscillations can be derived from Eqs. (\ref{12}) and
(\ref{14}). Let us suppose that the linear acceleration is zero, $\vec{a}=0$, and the
reference frame is rotating. In this situation, one can define a critical angular
velocity $\omega_c$ such that the off--diagonal matrix elements of (\ref{12}) vanish
 \beq\label{omegac}
 \Delta\sin 2\theta\approx \frac{2\hbar}{c}\,\vec{\omega}_c\cdot
 \vec{p}\,{,}
 \eeq
 implying that $\tilde\theta$ can assume the values $\tilde\theta \approx 0$
and $\tilde\theta \approx \pi/2$. The consequences of these values are discussed below
and only experimental tests may distinguish between them.
 For ultrarelativistic neutrinos, $E_{\nu}\sim pc$, it reduces to
 \beq\label{m1-m2}
 \vert m_2^2-m_1^2\vert \approx \frac{4\hbar E_{\nu}\omega_c}{\sin
 2\theta}\,{.}
 \eeq
 This formula connects the mass--squared difference of neutrinos to the
vacuum mixing angle, the neutrino energy  and the (critical) angular velocity of the
reference frame. In order to infer some consequences from (\ref{m1-m2}) for solar
neutrino problem (vacuum oscillations), we assume the reference frame co-moving with the
Earth, i.e. its angular velocity is $\omega_c\sim 7\cdot 10^{-5}$rad/sec. Results for
typical values of the neutrino energies and vacuum mixing angle are reported in Table I.
The agreement with the experimental data comes from neutrinos with energy varying in the
range $10\div 60$MeV. In this range, we find a mass--squared difference of the order
$10^{-12}\div 10^{-10}$eV$^2$ for vacuum mixing angle $10^{-1}\leq \sin 2\theta \leq 1$.
We have also considered the lower limit $10^{-1}$ for including, in such a way, the
actual uncertainty on the values of the vacuum mixing angle and show that, in any case,
our results are in very good agreement for $\sin 2\theta$ belonging to the range
$0.1\div 1$. Besides, we observe that an extreme value of $\tilde{\theta}$ as function
of $\theta$ (Eq. (\ref{14})) is
 \beq\label{tilth}
 \tilde{\theta}=\theta+\frac{\pi}{4}\,{.}
 \eeq
 On the other hand, the condition (\ref{omegac}) implies
$\tilde{\theta}\approx 0$ or $\tilde{\theta}\approx \pi/2$, which fix the vacuum mixing
angle approximatively to $\theta\approx \pi/4$, as expected by experimental results for
solar neutrinos.

The value $\tilde{\theta} \approx 0$ implies that, for a rotating observer, no mixing
occurs since in the flavor oscillations, due to the non--zero mass of neutrinos and to
the vacuum mixing angle, the term $\Delta\sin 2\theta$ is compensated by the geometric
term $p\cdot A_G$. As a consequence, the neutrino flux is different by the expected one
if inertial effects are neglected.

The value $\tilde{\theta}\approx \pi/2$ induces a conversion phenomena due to which the
flux of $\nu_e$ component decreases. To be more specific, after the production of
electronic neutrinos in the Sun, we have $\vert \psi (0)>=\vert \nu_e>$ ($\theta_0=0$).
Evolving along its worldline, the $\nu_e$ component will oscillate in agreement to Eqs.
(\ref{13}). Nevertheless, if condition (\ref{14}) holds and $ \tilde{\theta}\approx
\pi/2$, the probability (\ref{19}) to find $\nu_e$ in the beam decreases from 1 to
$\sin^2\theta_0\approx 0$. This result shows that the $\nu_e$ component of the beam is
almost totally depleted with respect to the rotating observer with angular velocity
$\omega_c$, resulting in a reduction of solar neutrino flux.

Concerning the atmospheric neutrino oscillation, an appreciable value of $\vert
m_2^2-m_1^2\vert$ requires highly (multiGeV) energetic neutrinos. In fact,  Eq.
(\ref{m1-m2}) implies $\vert m_2^2-m_1^2\vert\sim 10^{-4}$eV$^2$, for $E_{\nu}\sim
10^4$GeV and $\sin 2\theta\sim 10^{-2}$. This value of the mixing angle is excluded (at
least till now) by experimental data. Using $\sin^2 2\theta \geq 0.82$, according to the
experimental results, Eq. (\ref{m1-m2}) leads to a mass--squared difference of the order
$10^{-6}$eV$^2$, which does not fit the experimental range $10^{-4}\div 10^{-3}$eV$^2$.

In the regime in which neutrinos are highly energetic so that the condition
$\vec{p}\cdot\vec{A}_G\gg \Delta\sin 2\theta$ holds, Eq. (\ref{14}) implies
$\tilde{\theta}\approx \pi/4$, and the probability to find $\nu_e$ component in the
neutrino beam is $\approx 1/2$, assuming as initial condition $\vert\psi
(0)>=\vert\nu_e>$, $\theta_0=0$. Highly energetic neutrinos, with energy of the order
$1\div 10^3$TeV, can be produced, for example, by a Supernova \cite{PIN}.

It is interesting to compare the contribution to neutrino oscillations due to the
rotation term with that one due to the massive term. Being $p\cdot
A_G=-(\sqrt{-g}/c)\vec{\omega} \cdot\vec{p}$, Eq. (\ref{7}) implies that, for
ultrarelativistic neutrinos, the angular velocity of the rotating frame is given by
 \beq\label{24}
 \omega\sim \frac{1}{2}\frac{m_{\nu}^2c^4}{\hbar E_{\nu}}\,{.}
 \eeq
 If we consider neutrinos with energy $E_{\nu}\sim 1$ TeV,
emanating from active galactic nuclei \cite{PIN} which are possible sources of high
energy signal, being the most luminous objects in the Universe \cite{BER}, and for
$m_{\nu}\sim 1$ eV$/c^2$ \cite{PIN},\cite{GGR}, the angular velocity is of the order
$\omega\sim 10^2\div 10^3$ rad/sec. Moreover, one aspects that the neutrino mass is of
the order $m_{\nu}\sim 10^{-2}\div 10^{-4}$ eV$/c^2$ \cite{ALT},\cite{HAX}. At energy
$E_{\nu}\sim 10\div 10^2$ GeV, produced by accelerators \cite{NJB}, one gets $\omega\sim
1.2\div 10^{-5}$ rad/sec.

Some values of the angular velocity, calculated by using Eq. (\ref{24}) for different
(and expected) values of neutrino masses and energies, are reported in Table II. It
turns out that these values are of the same order of magnitude of typical angular
velocity of astrophysical objects (Table III). For example, the angular velocity
$\omega\sim 10^2\div 10^3$rad/sec is comparable with that one of pulsars \cite{HAR}.
Besides, $\omega\sim 10^{-5}$rad/sec is of the same order of angular velocity of the
Earth, and it is about $10$ times the angular velocity of the Sun, $\omega_{Sun}\sim
10^{-6}$rad/sec. An experiment (as for example, an interferometer experiment) aimed to
measure the quantum mechanical phase shift induced by such a geometric term is difficult
to realize since the detector must be comoving with the astrophysical objects. A
sensible effect could be tested for Earth assuming the above values of $m\sim
10^{-2}\div 10^{-4}$eV/c$^2$ and $E\sim 10\div 10^2$GeV, but it requires a so high
precision that, at the moment, it is not allowed by technology.

In the case in which  the acceleration $\vec{a}$ is constant and $\vec{\omega} =0$, the
$p\cdot A_G$ term in Eq. (\ref{7}) vanishes and one finds a shift of the phase:
$\Omega=i\int_{\lambda_0}^{\lambda}P\cdot p_{null}d\lambda^{\prime}$, as defined in Eq.
(\ref{1}). In fact, since the neutrino trajectory is null and $g_{\mu\nu}$ is diagonal
($\sqrt{-g}=1+(\vec{a}\cdot\vec{x})/c^2$), the physical distance can be written as
 \beq\label{20}
 d\lambda = dl\left(g_{ij}\frac{dx^i}{d\lambda}\frac{dx^j}{d\lambda}\right)^{-1/2}=
dl\left[-g_{00}\left(\frac{dx^0}{d\lambda}\right)^2\right]^{-1/2}\,{,}
 \eeq
 and one gets \cite{CAR}
 \beq\label{21}
\Omega=-\frac{M^2}{2E_*}\int_{l_0}^{l}\frac{1}{1+(\vec{a}\cdot\vec{x})/c^2}dl^{\prime}\,{,}
 \eeq
 where $E_*=P_t$ is the conserved quantity due to the non
dependence of the metric tensor on the timelike coordinates. If
$\vec{a}\parallel\vec{x}$, $dl=dr$ and Eq. (\ref{21}) reduces to
 $$ \Omega=-\frac{M^2c^2}{2E_* \vert\vec{a}\vert
}\ln(1+\vert\vec{a}\vert r/c^2)\,{,}
 $$
 for $r_0\equiv r(l_0=0)=0$. If $\vec{a}\perp\vec{x}$, Eq. (\ref{21}) gives the
standard result $\Omega=-(M^2c^2/(2E_*)(\lambda-\lambda_0)$.

\section{Conclusions}

We have analyzed the phenomenological aspects of neutrino oscillations for an
accelerating and rotating observer.

The inertial effects on neutrino oscillations seem to be appreciable on solar neutrino
problem. The coupling of the angular velocity and momentum of the neutrino implies a
reduction of neutrino flux, as experimented by an observer comoving with the Earth,
providing us with valuable informations on mass--squared difference and mixing angle of
neutrinos. In fact, we find a good agreement between the experimental data \cite{ALL1}
and our estimations of the mass--squared differences. They well fit the experimental
data for neutrino beams with energies of the order $10\div 60$MeV. For these estimations
we have used $\sin 2\theta\sim 1\div 10^{-1}$, values coming from the data of solar
neutrino oscillation experiments.

In the framework of atmospheric neutrinos, inertial effects seems to be negligible. The
best--fit of experimental data that we reproduce, if adiabatic condition holds, comes
for highly energetic neutrinos. We get in fact, a value of the mass--squared difference
of the order $10^{-4}$eV$^2$, requiring a  mixing angle $\sin 2\theta\sim 10^{-2}$.

However, we have to note that values of the mass--squared difference of neutrinos and
their mixing angle are till now open issues. Only future neutrino oscillation
experiments will make possible to investigate in detail the region $\vert
m_2^2-m_1^2\vert$ and to fix the value of the mixing angle. These data will allow to
solve definitely the solar and atmospheric neutrino problems and to understand if
inertial effects are important for explaining the deficit of solar and atmospheric
neutrino flux.

\vspace{2cm}

\centerline{\bf Acknowledgment}

The authors would like to thank the referee for the useful comments which allowed to
improved the paper.

\newpage
\begin{center}
Table I: Estimation of $\vert m_2^2-m_1^2\vert$ as function of $E_{\nu}$, $\sin 2\theta
$ and fixed value of $\omega_c$.
\end{center}
\begin{center}
\begin{tabular}{|c|c|c|} \hline\hline
$E_{\nu}$(MeV) & $\sin 2\theta$ & $\vert m_2^2-m_1^2\vert$(eV$^2$)
\\ \hline\hline
 1 & 1 & $10^{-13}$ \\
 1 & $10^{-1}$ & $10^{-12}$ \\ \hline\hline
 10 & 1 & $10^{-12}$ \\
 10 & $10^{-1}$ & $10^{-11}$ \\ \hline\hline
 $50\div 60$ & 1 & $10^{-10}$ \\ \hline\hline
 \end{tabular}
 \end{center}

\bigskip

\begin{center}
Table II: Angular velocity of reference frames for different values of neutrino masses
and energies.
\end{center}
\begin{center}
\begin{tabular}{|c|c|c|} \hline\hline
$m_{\nu}$(eV/c$^2$)             & $E_{\nu}$(GeV)       & $\omega$(rad/sec) \\
\hline\hline
   1       & $10^3\div 10^{-1}$ & $10^2\div 10^6$  \\ \hline
 $10^{-2}$ & $10^3\div 10^{-1}$ & $10^{-2}\div 10^2$   \\ \hline
 $10^{-4}$ & $10^3\div 10^{-1}$ & $10^{-6}\div 10^{-2}$ \\ \hline\hline
\end{tabular}
\end{center}

\bigskip

\begin{center}
Table III: Typical angular velocity of Astrophysical objects.
\end{center}

\begin{center}
\begin{tabular}{|c|c|} \hline\hline
 Astrophysical Objects       & $\omega$(rad/sec)\cite{HAR} \\ \hline\hline
 Pulsar                      &  $4\cdot 10^3$      \\
 Sun                         &  $10^{-6}$     \\
 Earth                       &  $10^{-5}$     \\
 White dwarf                 &  $ 2.1$         \\
 RRLyrae Star                &  $10^{-5}$          \\
 Cepheid Variable            &  $ 10^{-7}$         \\ \hline\hline
\end{tabular}
\end{center}

\end{document}